\documentclass[preprint,12pt]{elsarticle}

\def\s1{\hat{s}_1}
\def\T1{\hat{t}_1}
\def\t2{\hat{t}_2}
\def\U1{\hat{u}_1}
\def\u2{\hat{u}_2}

\newcommand{\comment}[1]{}

\usepackage{amssymb}

\journal{Physics Letters B}

\begin{document}

\begin{frontmatter}



\title{ {\bf Probing the spin-parity of the Higgs boson 
via  jet kinematics in  vector boson fusion}}

\author[abdel1]{A. Djouadi}
\author[iisc]{R.M. Godbole}
\author[bruce1,bruce2]{B. Mellado}
\author[iisc]{and K. Mohan} 

\address[abdel1]{LPT, Universit\'e Paris-Sud and CNRS, 91405 Orsay Cedex, France}
\address[iisc]{Centre for High Energy Physics, Indian Institute of 
Science, Bangalore 560 012, India.}
\address[bruce1]{Department of Physics, University of Wisconsin, 
Madison, WI 53706, USA. }
\address[bruce2]{University of the Witwatersrand, School of Physics,
Private Bag 3, Wits 2050, Johannesburg, South Africa. }


\begin{abstract}

Determining  the spin and the parity quantum numbers of the  recently discovered
Higgs--like boson at the LHC is a matter of great importance.  In this paper, we
consider the possibility of using the kinematics of the tagging jets in Higgs
production via the vector boson fusion (VBF) process to test the tensor
structure of the Higgs--vector boson ($HVV$) interaction and to determine the
spin and CP properties of the observed resonance. We show that an anomalous
$HVV$ vertex, in particular its explicit momentum dependence, drastically
affects  the rapidity between the two scattered quarks and their transverse
momenta and, hence, the acceptance of the  kinematical cuts that allow to select
the  VBF topology. The sensitivity of these observables to different spin-parity
assignments, including the dependence on the LHC center of mass energy, are
evaluated.  In addition, we show that in associated Higgs production with a
vector boson some kinematical variables, such as the invariant mass of the 
system and the transverse momenta of the two bosons and their separation in
rapidity, are also sensitive  to the spin--parity assignments of the Higgs--like
boson.   \end{abstract}

\comment{
}

\end{frontmatter}


\section{Introduction}

The mechanism for electroweak symmetry breaking~\cite{EWSB} is minimally
realized in the  Standard Model (SM) of particle physics~\cite{GSW} by
introducing a  doublet of complex scalar fields that develops a non-zero vacuum
expectation value;  this leads to  the existence of  only one physical scalar
boson $H$, with  the spin, parity and charge conjugation assignments of a
$J^{\rm PC}\! =\! 0^{++}$ state. 

Accumulated data by the ATLAS and CMS collaborations seem to
indicate that the newly observed bosonic particle  with a mass around 125 GeV
\cite{Higgs_search} has production and decay rates that are compatible with
those expected for  the SM Higgs particle. However, it will  be necessary that experiments 
test the spin and CP properties of the new boson before
we can truly identify the observed particle with the Higgs boson of the minimal
SM. In particular, more data will be required in order to verify  that the Higgs
sector is not extended in order to contain CP--violating interactions that could
make the observed Higgs particle a mixture of CP--even and CP--odd states.  In
addition, while it is clear from the observation of the $\gamma\gamma$  decay
channel that  the observed bosonic state cannot have spin equal to one
\cite{Landau-Yang},  it is not yet entirely excluded that it has a spin equal to
two or more.

It is well known that the structure of the Higgs to vector boson ($HVV$)
coupling is a possible tool to probe the Higgs $J^{\rm PC}$ quantum 
numbers\footnote{Strictly speaking a $0^{-+}$ CP--odd state will not have a
tree--level coupling to massive gauge bosons;  the coupling is generated 
through loop corrections and should be thus small. We will nevertheless,
for the sake of comparison,  follow the current trend and allow for couplings to
VV states that are of the same order as those of the SM $0^{++}$ CP--even
state.}  at the LHC , ILC and the LHeC. Distributions in various kinematic variables,
the virtuality of the off--shell gauge boson, and angular correlations of the
decay products in the decay channel $H\to VV \to 4f$ with $V=W,Z$ in particular,
allow to discriminate between CP--even and CP--odd states as well as spin--zero
from higher spin particles
\cite{CP-papers,ilc-hvv-ref,Biswal:2012mp,Choi:2002jk,reviews,Hagiwara:2009wt}.
It is also known that the  azimuthal angle distribution of the two outgoing
forward tagging jets in  Higgs production in the  vector boson fusion (VBF)
process effectively discriminates between the CP quantum numbers of a scalar
resonance~\cite{PleRaiZep,azimuth} as well as the spin
\citep{Hagiwara:2009wt}.    

In the present paper, we consider Higgs production in the vector boson fusion
mechanism,  $pp \to Hjj$, in the presence of an anomalous Higgs--vector boson
vertex that parametrises different spin and CP assignments of the produced 
state. The anomalous $HVV$ coupling is introduced by allowing for  an effective
Lagrangian with higher dimensional operators, that include four momentum terms 
which are absent in the SM.  We show that the kinematics of the forward tagging
jets in this process is highly sensitive to the structure of the anomalous
$HVV$  coupling and that it can effectively discriminate between different
assignments for the spin (spin-0 versus spin-2) and the parity  (CP--even versus
CP--odd)  of the produced  particle.  

We find, in particular, that the correlation between the separation in rapidity
and the transverse momenta of the scattered quarks in the VBF process, in
addition  to the already discussed distribution of the azimuthal jet
separation,  can be significantly altered compared to the SM expectation. These
kinematical variables define new corners of the phase-space that have not been
explored by the experiments at the LHC to  probe anomalous $HVV$ couplings and
check the  $J^{\rm PC}$ assignments of the newly observed particle. Some of
these observables significantly depend on the center of mass energy  and strong
constraints on anomalous couplings can be obtained by performing measurements at
the LHC with energies of $\sqrt s = 8$ TeV and $\sqrt s = 13$ TeV.  Finally, we
also consider  associated Higgs production with a massive gauge boson (VH),
$q\bar q \to VH$ with $V=W^\pm$ and $Z$, and show that the invariant mass of the
$VH$ system as well as the transverse momenta and rapidities of the $H$ and $V$
bosons are also sensitive to anomalous $HVV$ couplings.

An analysis that is, in some aspects, similar to ours has been recently reported
in Ref.~\cite{Englert:2012xt} and we will discuss the main differences between 
this study and the one  presented here  at the end of the paper. 

For the rest of this article, the next section  describes the physical set--up
and  sections~\ref{sec:vbf} and \ref{sec:vh} our phenomenological analyses  in
the VBF and VH processes;  section~\ref{sec:conclusions}
summarizes the main findings.

\section{The physical set-up}
\label{sec:effla}

In the SM, the couplings of the Higgs boson to the massive electroweak gauge 
bosons are precisely formulated and come out as  $g_{HVV}
\propto  g M_V V_\mu V^\mu$ with $g$ the SU(2) coupling constant.  However, this
is not the most general form of the Higgs--gauge boson vertex. Parametrising the
coupling of a scalar state to  two vector bosons  in the form $i\Gamma^{\mu\nu}
(p,q) \epsilon_\mu(p) \epsilon^\ast_\nu(q)$, one can write down the most general
form of the $HVV$  vertex as $\Gamma_{\mu\nu}(p,q)= \Gamma^{\rm  SM}_{\mu\nu}
+\Gamma^{\rm BSM}_{\mu\nu}(p,q)$, with the SM and the beyond SM  components
given by
\begin{eqnarray}
\Gamma^{\rm  SM}_{\mu\nu} &=& - gM_V\, g_{\mu\nu} \label{eqn:SMvertex} \\
\Gamma^{\rm BSM}_{\mu\nu}(p,q) &=& \frac{g}{M_V}\left[ 
\lambda \left(p \cdot q\, g_{\mu\nu} - p_\nu q_\mu  \right)
+ \, \lambda^\prime\ \epsilon_{\mu\nu\rho\sigma}p^\rho q^\sigma \right]
\label{eqn:BSMvertex}
\end{eqnarray}
where $\lambda$ and $\lambda^\prime$ are effective  coupling strengths for,
respectively, higher dimension CP--even and CP--odd operators and
we will assume that they are the same for $W$ and $Z$ bosons. These operators
may be generated within the SM at higher orders of perturbation theory, 
although the resulting couplings are likely to be very small. In
general, $\lambda$ and $\lambda^\prime$ can be treated as momentum dependent
form factors that may also be complex valued. However, we take the approach
that  beyond the SM (BSM) vertices can be generated from an effective
Lagrangian, which treats $\lambda$ and $\lambda^\prime$ as coupling
constants~\cite{PleRaiZep}.  The most striking difference between the SM and BSM
vertices of eqs.~(\ref{eqn:SMvertex}) and (\ref{eqn:BSMvertex}) is that the 
latter has an explicit dependence on the momentum of the gauge bosons. It is
this feature that is the source of the differences that the  BSM vertices
generate in the kinematic distributions of tagging jets in the VBF and VH
processes, compared to the SM case.

In the case of a spin-2 resonance coupling to $VV$ states, we will follow 
Ref.~\cite{j-frank}  and adopt the  following effective Lagrangian 
\begin{equation}
 \mathcal{L}_{\rm eff} =\frac{1}{\Lambda} T_{\alpha \beta} \; 
 (f_1 g_{\mu\nu}B^{\mu \alpha} B ^{\nu \beta}  + 
  f_2 g_{\mu\nu}W^{\mu \alpha} W ^{\nu \beta})
 \label{eqn:spin2-lag}
\end{equation}
where $T_{\alpha \beta}$ is the spin-2 field and  $B_{\mu \nu}, W_{\mu \nu}$ 
are the U(1) and SU(2) field strengths, respectively, and $\Lambda$ corresponds
to a cut-off scale which should be set at a value of the order of the TeV
scale. The two operators in the equation above do not  account for all  possible
terms of  the most general spin-2--VV vertex and one could, for instance,
include  operators such as  $\partial_\mu  \partial_\nu T_{\alpha  \beta}(
\tilde{B}^{\mu \alpha} B ^{\nu \beta}\! + \!  \tilde{W}^{\mu \alpha} W ^{\nu 
\beta})$ but which are of higher dimension. We restrict ourselves to the terms
proportional to $f_1$ and $f_2$ of eq.~(\ref{eqn:spin2-lag}), setting the
cut--off scale to $\Lambda=3$ TeV in our numerical illustration
.

The purpose of our analysis  is to identify observables and model independent
tendencies in the kinematics of partons due to the anomalous $HVV$ couplings 
above in electroweak processes   in which the Higgs--like boson is produced in
association with two jets, i.e. in vector boson fusion   and in associated Higgs
production  with vector bosons.  For a given choice of the spin and parity
assignments, we  identify regions of the phase--space that would be populated
differently by the Higgs--like boson with anomalous couplings  compared to the
expectation in the SM.

In our analysis, the vertices for the Lagrangians in the SM and in BSM with
spin-0 and spin-2 bosons are calculated  in {\tt FEYNRULES}~\cite{FEYNRULES} and
passed to the program {\tt Madgraph}~\cite{Alwall:2011uj}, which is used for the
generation of  the matrix elements for Higgs production in VBF and VH. To
obtain the cross sections and distributions at the hadronic level, the CTEQ6L1
parton distribution functions  are used~\cite{CTEQ}. For completeness,
associated SM Higgs production with two high $p_T$ jets in  the gluon--gluon
fusion process (ggF) is also considered;  the cross sections and kinematical
distributions are obtained using the MCFM~\cite{ggF2jNLO} program that
incorporates the  QCD process $pp\rightarrow H+2j+X$ at next-to-leading order
(NLO). The factorization and normalization scales are set on an event-by-event
basis to the transverse energy of the Higgs boson. 
For the selection cuts, 
partons are required to have
transverse momentum $p_T>20\,$GeV, rapidity $\left|y\right|<4.5$ and be separated
by $\Delta R>0.7$. 

\section{Spin and parity determination in  VBF}
\label{sec:vbf}

The differences in the tensor structure of the different terms in the $HVV$
vertex would significantly impact the kinematic distributions of the tagging
jets produced in VBF. For example, in
Ref.~\cite{PleRaiZep} the difference in azimuthal angle between the scattered
quarks ($\Delta\phi_{jj}$)  was found to be an effective discriminant for the
various terms in eq.~(\ref{eqn:BSMvertex}) for a scalar boson. As we will show 
below,  the momentum dependence of the BSM vertex has further strong
implications for a number of hadronic observables. These include the transverse
momenta of the tagging partons\footnote{Tagging partons are defined as the
scattered quarks in  the process $pp\to Hjj$; at NLO, these are defined as the
leading partons in transverse momentum.} ($p_{Tj1},  p_{Tj.2}$) and the rapidity
difference ($\Delta y_{jj}$) between them. The obvious advantage of looking at
such observables is that they do not require a full reconstruction of the
Higgs--like boson from its decay products and one can use all the search
channels at our disposal at the LHC, including the $H\to \gamma \gamma$ and 
$H\to \tau^+\tau^-$ modes. 

In this section we describe the effect on the rapidity and transverse
momentum distributions of the jets due to each of the operators of eqs.~(2--3),
without considering any interference between them. In doing so it is important
to understand that the distributions are independent of the strength or sign of
the couplings $(\lambda,\lambda^\prime,f_1,f_2)$ which only affect the value of
the total cross-section. The strength of these couplings play an important role
when one considers the simultaneous presence of these operators in the $HVV$
vertex and we will discuss this at the end of this section.

\begin{figure*}[t]
\begin{center}
\includegraphics[height=6cm]{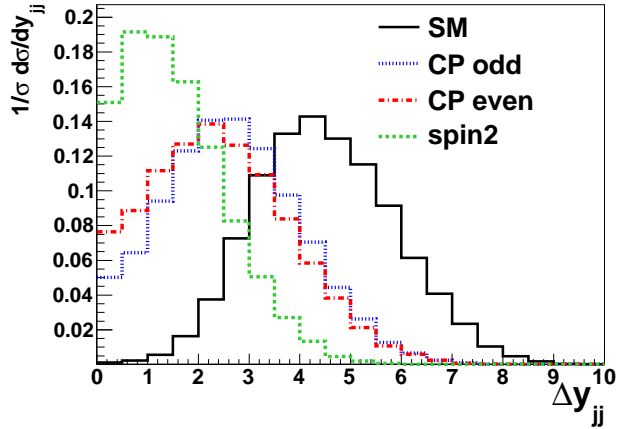}\\ 
\end{center}
\vspace*{-3mm}
\caption[]{Normalized distribution of the difference in rapidity between the 
scattered quarks in the VBF process for  each of the SM (black-solid-line) and 
BSM operators individually; the spin2 (green-dashed)  CP even (red-dotted-dashed) and the CP odd (blue-dotted).}
\label{fig:vbf1Da}
\vspace*{-1mm}
\end{figure*}

Figure~\ref{fig:vbf1Da} displays the difference in rapidity between tagging
partons ($\Delta y_{jj}$) for each of the higher dimensional operators.

We see that the rapidity differences for each of the operators 
are shifted to smaller values. While the BSM $0^+$ and $0^-$ cases display a behavior in this observable that is almost identical, for a spin-2 particle the peak in the $\Delta y_{jj}$ distribution is shifted to even smaller values.

The dramatic behavior of this observable makes it a strong candidate to distinguish between the SM and the BSM vertex structures.

We also observe that the transverse momenta of the tagging partons become significantly larger
for pure BSM $0^+$ and $0^-$ states. This feature is more pronounced for a 
spin-2 state. However,  the $M_{jj}$ distributions are not altered and we 
will thus not display them.

The above mentioned features of the $\Delta y_{jj}$ and $p_T$ distributions 
are due to the presence of momentum dependent structures in the
vertices of eqs.~(\ref{eqn:BSMvertex},\ref{eqn:spin2-lag}). 

The factors involving four momenta in the BSM vertices mean that the momentum
flow through the vertex allows for a greater push being given to the jets
resulting in a shift of the $p_T$ distribution to larger values. This effect has
been observed in~Refs.~\cite{zep-figy,j-frank} for both the scalar and the
spin-2  case. 

An analytic calculation of the matrix element reveals that for
both the SM and BSM cases, the vector boson propagator can be approximated to
$1/(\hat{s}\; p_{Tj1}\; p_{Tj2} e^{-\Delta y_{jj}})^2$ for large values
of the incoming parton momenta. The effect of this term is to push $p_{Tj}$ to
smaller values and, at the same time, push $\Delta y_{jj}$ to larger values. 
For the SM, the remainder of the terms in the square of the matrix element are
proportional to $\hat{s}\; p_{Tj1}\; p_{Tj2} \; {\rm Cosh}(\Delta y_{jj})$, which 
further reinforces the  occurrence of large
rapidity differences. 

For  $\Gamma^{\rm BSM}_{\mu\nu}$ of eq.~(\ref{eqn:BSMvertex}), the additional
terms have a dependence of the form $(\hat{s}\; p_{Tj1}\; p_{Tj2})^2$. This
leads to much larger and flatter $p_{Tj}$  distributions as compared to the SM.
Although the rapidity dependence  for these additional terms  is complicated, it can be
shown to have an opposite behavior to the propagator terms, unlike the SM case,
pushing the rapidity difference to  smaller values.  Thus, the correlations
between $\Delta y_{jj}$ and $p_{Tj1}$, $p_{Tj2}$ and/or $M_{jj}$ critically 
depends on the tensor structure of the $HVV$ vertex.

\begin{figure*}[!t]
\begin{center}
\hspace{-6mm}
\includegraphics[height=5cm]{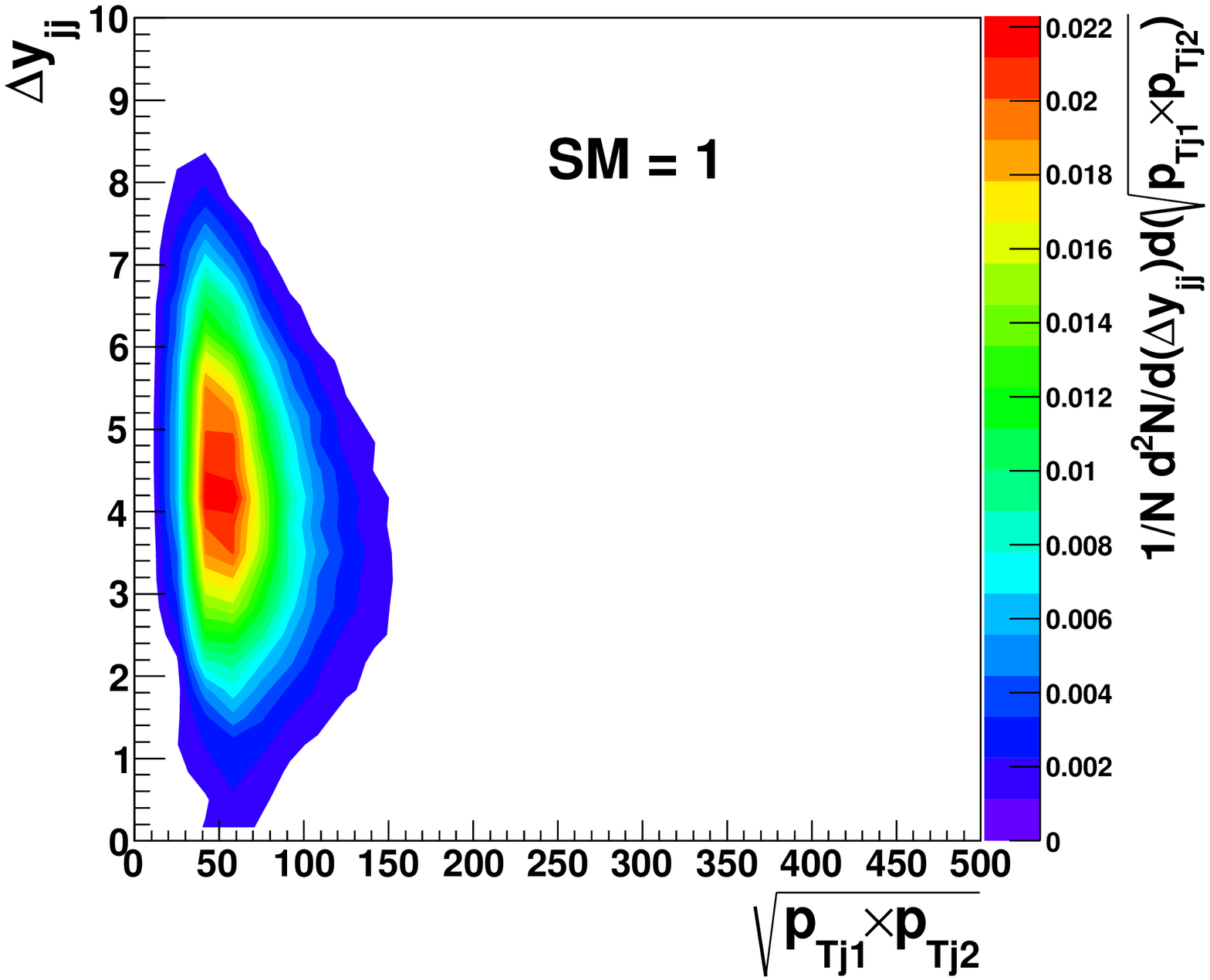}~~~ 
\hspace{-5mm}
\includegraphics[height=5cm]{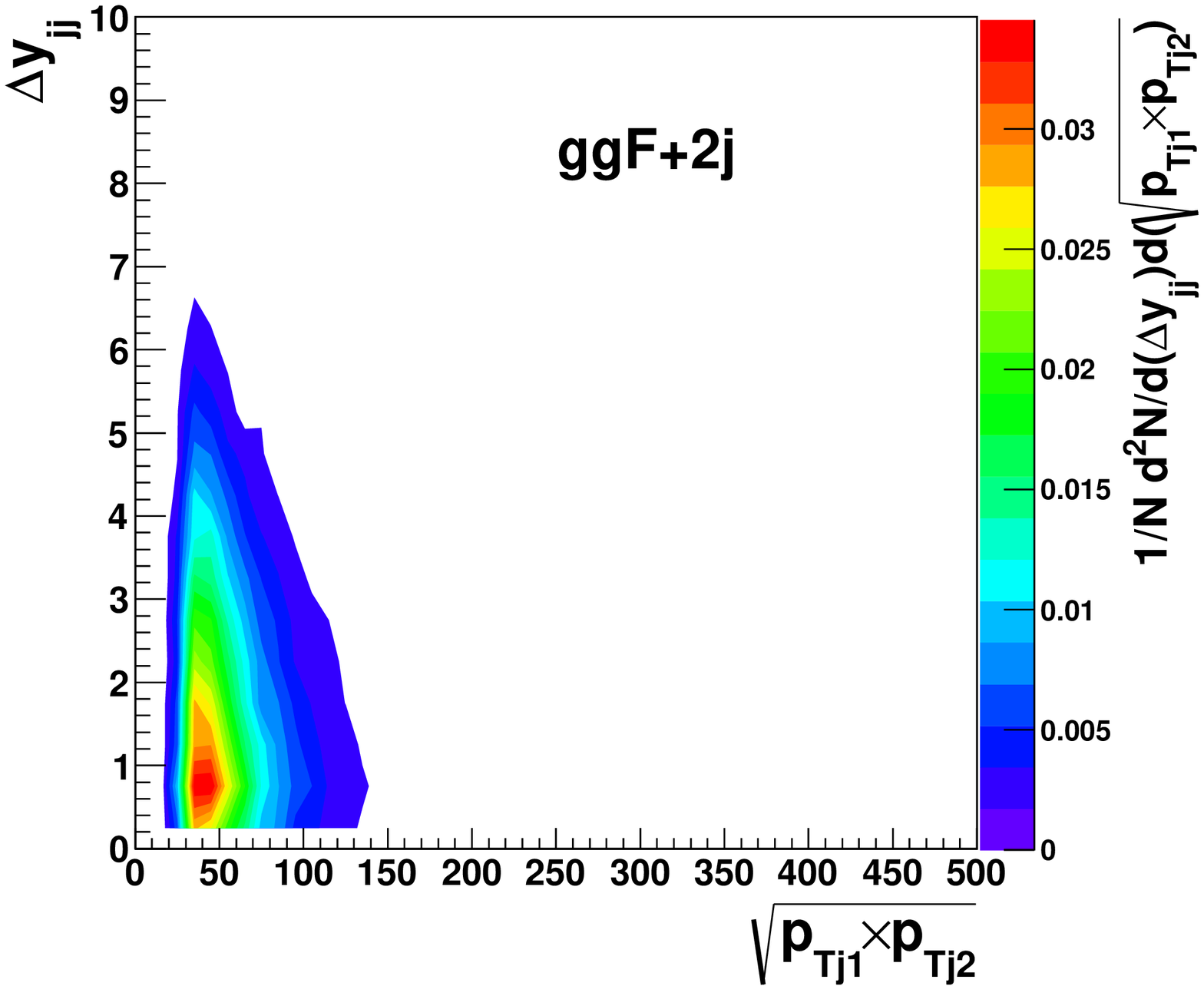}
\\[2mm]
\includegraphics[height=5cm]{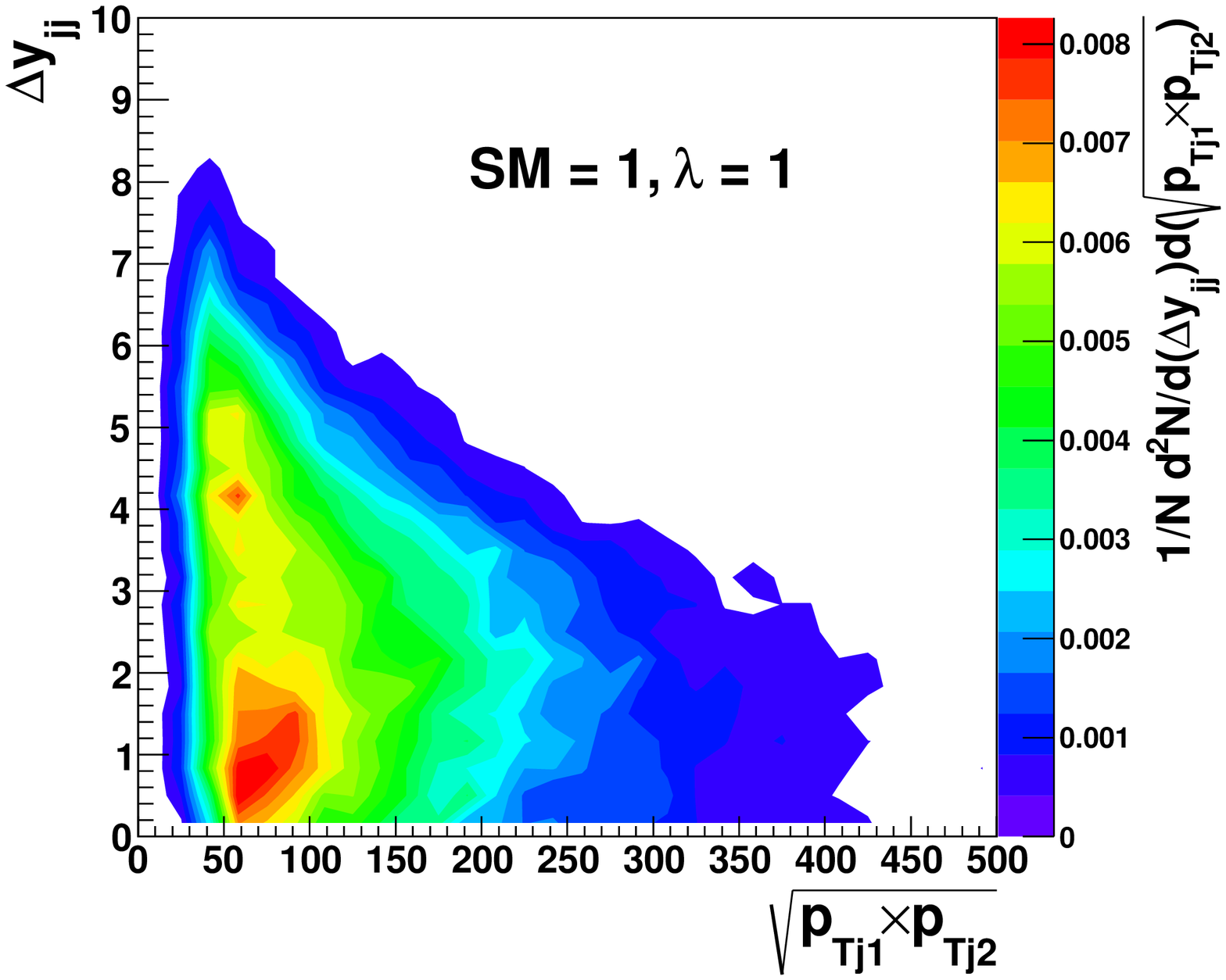}
\includegraphics[height=5cm]{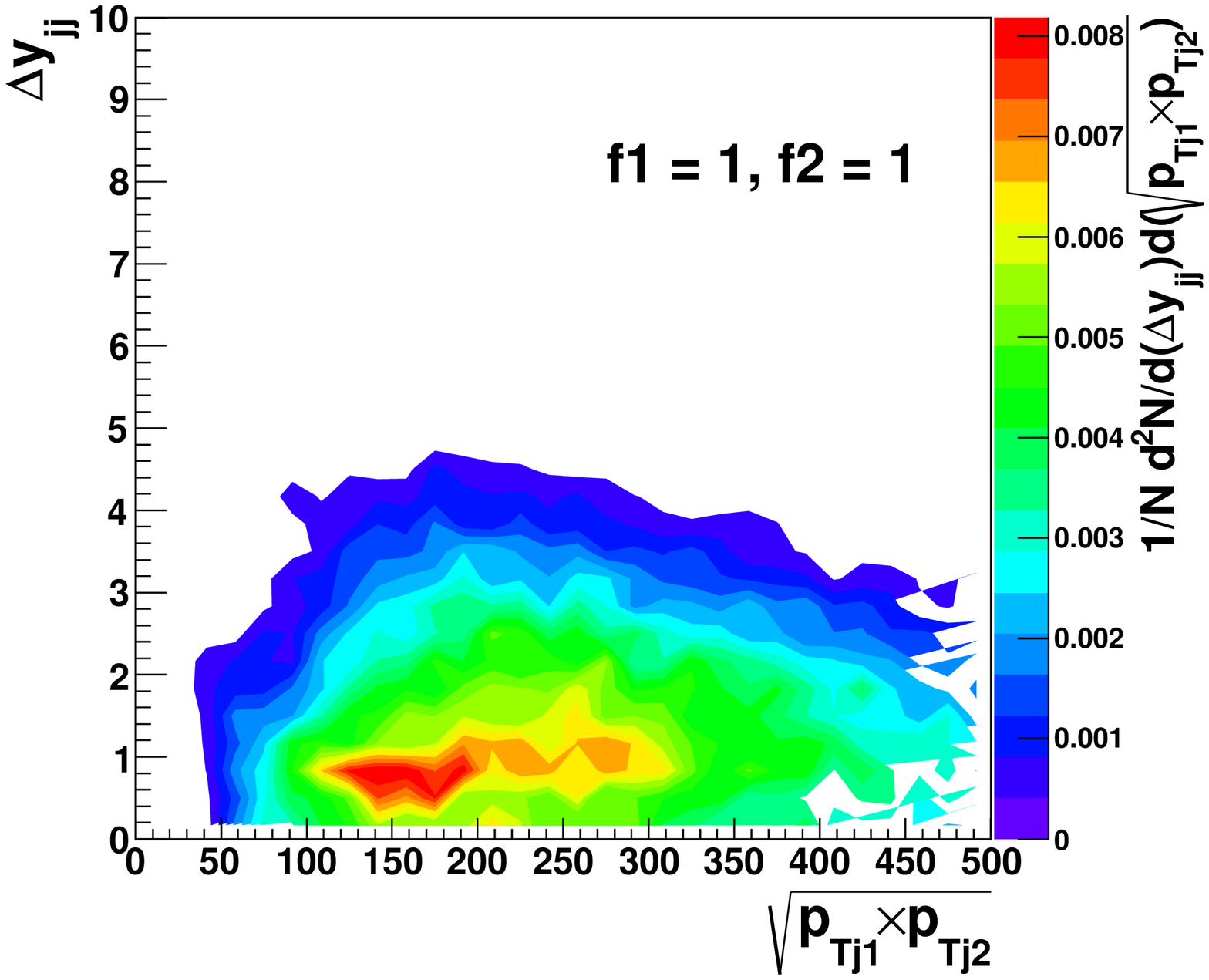}~~~
\vspace*{-3mm}
\end{center}
\caption[]{Two-dimensional distributions of the difference in rapidity and the 
square root of the product of the transverse momenta of the tagging partons. 
The upper plots correspond to the SM with electroweak (left) and ggF$+2j$ 
(right) production while the lower plots to an admixture with a BSM $0^+$ 
state with $\lambda=1$ (left) and a pure $2^+$state (right). Points with 
the same cross-section are indicated with color contours.
The level of cross-sections is indicated by the color code in the bar on the right of each of the plots. Thus area with the largest and lowest cross-sections is indicated with dark red and dark blue respectively.}
\label{fig:vbf2D}
\vspace*{-2mm}
\end{figure*}

This is further illustrated in Fig.~\ref{fig:vbf2D}, where we  use the
observable $\sqrt{p_{Tj1}\cdot p_{Tj2}}$ in view of the above
considerations. This figure also suggests a new region of the phase-space for
the exploration of physics beyond the SM in the Higgs  sector by using
hadronic observables. BSM vertices tend to populate a region of the phase-space
where the production of the SM Higgs boson is significantly depleted.  Three
distinct regions of the phase-space in the Higgs boson production in association
with two highest $p_T$  partons are identified:

\begin{itemize}

\item { The QCD region} in which the production of ggF+$2j$ is dominant. Here, 
the partons tend to be close in rapidity and display moderate $p_T$ with 
$<\sqrt{p_{Tj1}\cdot p_{Tj2}}>\approx 50$ GeV. In this corner of the 
phase-space, associated Higgs production with a vector boson with $V\rightarrow
jj$ contributes as a sub-leading process for which $<\sqrt{p_{Tj1}\cdot
p_{Tj2}}>\approx 45$ GeV.

\item { The SM VBF region} which  is populated by the SM Higgs boson produced
via the VBF process. This corresponds to intermediate $p_T$ tagging partons with
$<\sqrt{p_{Tj1}\cdot p_{Tj2}}>\approx 55$ GeV with a large separation in
rapidity of $<\Delta y_{jj}> \approx 4.5$. Production of additional central
partons is depleted, which is a typical signature of electroweak processes. 

\item {The BSM electroweak region} which populates a corner of the phase-space
defined by\footnote{The separation in rapidity in the BSM electroweak region
tends to be somewhat larger compared to that displayed by ggF+$2j$ production.
For instance,  $<\Delta y_{jj}>=2.1$ for a pure BSM $0^+$ state compared to 
$<\Delta y_{jj}>=1.7$ for ggF+$2j$. } $\sqrt{p_{Tj1}\cdot p_{Tj2}}> 100$ GeV and
$\Delta y_{jj}<4$. The VBF and VH production channels can be separated by
requiring that the invariant mass of the leading partons be away from the weak
boson invariant mass. A distinct BSM VBF region is defined where
electroweak-like gluon radiation is expected to occur. This feature is important
to further differentiate and suppress QCD Higgs boson production.  

\end{itemize}

\begin{table}[t]
\renewcommand{\arraystretch}{1.3}
\begin{center} {\footnotesize
\begin{tabular}{c|cc||cc||cc||cc||c}
\hline
 & \multicolumn{2}{c||}{SM EW $Hjj$ } & \multicolumn{2}{c||}{BSM $0^+$} & \multicolumn{2}{c||}{$0^-$} & \multicolumn{2}{c||}{$2^+$} & SM ggF+$2j$ \\\hline
Process & VBF & VH & VBF & VH & VBF & VH & VBF & VH & \\
 Acceptance & 0.06  & 0.04 & 0.59 & 0.12 & 0.55 & 0.18 & 0.93 & 0.75 & 0.10 \\ 
 $\sigma$ (fb) & 0.14  & 0.04 & 1.43 & 0.29 & 1.35 & 0.43 & 2.25 & 0.86 & 0.27 \\  \hline
$\sigma^{EW}_{BSM}/\sigma^{EW}_{SM}$ & -  & - & 8.8 & & 8.7 & & 17.4 & &  - \\
\hline
 \end{tabular} }
\caption{Acceptance, expected cross-sections (in fb) and the ratio of BSM Higgs
boson cross-section to that  in the SM populating the BSM electroweak region 
($\Delta y_{jj}<4$ and $\sqrt{p_{Tj1}\cdot p_{Tj2}}>100\,$GeV). Pure BSM states
are considered.  The  cross-sections are given for the $H\rightarrow\gamma
\gamma$ decay and 8\,TeV center of mass energy (see text).}
\label{tab:BSMEW}
\end{center}
\vspace*{-6mm}
\end{table}
 
Table~\ref{tab:BSMEW} displays the acceptance of the SM Higgs boson and  other
BSM pure states in the BSM electroweak region. The latter is defined as $\Delta
y_{jj}<4$ and $\sqrt{p_{Tj1}\cdot p_{Tj2}}>100\,$GeV,  where tagging partons are
required to have $p_T>25\,$GeV and $\left|y\right|<5$.

The table shows the
acceptance of each process, separating the VBF and VH  mechanisms. 
The BSM cross-sections are assumed to be the same as the one for electroweak SM
$Hjj$ production which is normalized to the NNLO--QCD
value~\cite{Dittmaier:2011ti}; all cross sections  are given for the
$H\rightarrow\gamma\gamma$ decay.

The SM Higgs boson production displays acceptances\footnote{The acceptance of
the ggF process is reported with respect to the generator cuts given in
section~\ref{sec:effla}.}  of less than 10\%. In this process, additional
gluons are radiated more copiously than in the electroweak  processes. The
fraction of the events in the BSM electroweak region that survive a veto on
additional gluons with $p_T>25\,$GeV and $\left|y\right|<5$ is
$50^{+30}_{-25}\%$.   The last row gives the ratio of the BSM contribution to
that of the SM, where only electroweak processes are taken into account. In
order to give a rough estimate of the signal significance for different
scenarios with the $H\rightarrow\gamma\gamma$ decay, the non-resonant production
of two photons and two partons  is modeled with {\tt Madgraph}~\cite{Alwall:2011uj}.
The cross-sections of pure BSM signals, of background and the corresponding
signal significance are scanned as a function of the lower bound on
$\sqrt{p_{Tj1}\cdot p_{Tj2}}$.

It is found that for pure BSM scalar states maximal significance is obtained
with $\sqrt{p_{Tj1}\cdot p_{Tj2}}\gtrsim 100\,$GeV and for a pure spin-2 state
is obtained for $\sqrt{p_{Tj1}\cdot p_{Tj2}}\gtrsim 300\,$GeV. Assuming 8\,TeV
center of mass energy and 25\,fb$^{-1}$ of integrated luminosity, a significance
of approximately 1.4\,$\sigma$ and 5\,$\sigma$ can be achieved for the pure BSM
scalar and spin-2 cases considered here, respectively. This sensitivity is
achieved with signal-to-background ratios of 0.08 and 1.5 for the pure BSM
scalar and spin-2 states, respectively. This is defined in the di-photon
invariant mass window $120<M_{\gamma\gamma}<130\,$GeV.

It is important to note that the experiments at the LHC currently do not explore
the BSM electroweak region. This region is  vetoed by requiring a large
separation in rapidity of the tagging partons, in order to isolate the SM VBF
process.

\begin{figure*}[t]
\begin{center}
\includegraphics[height=6cm]{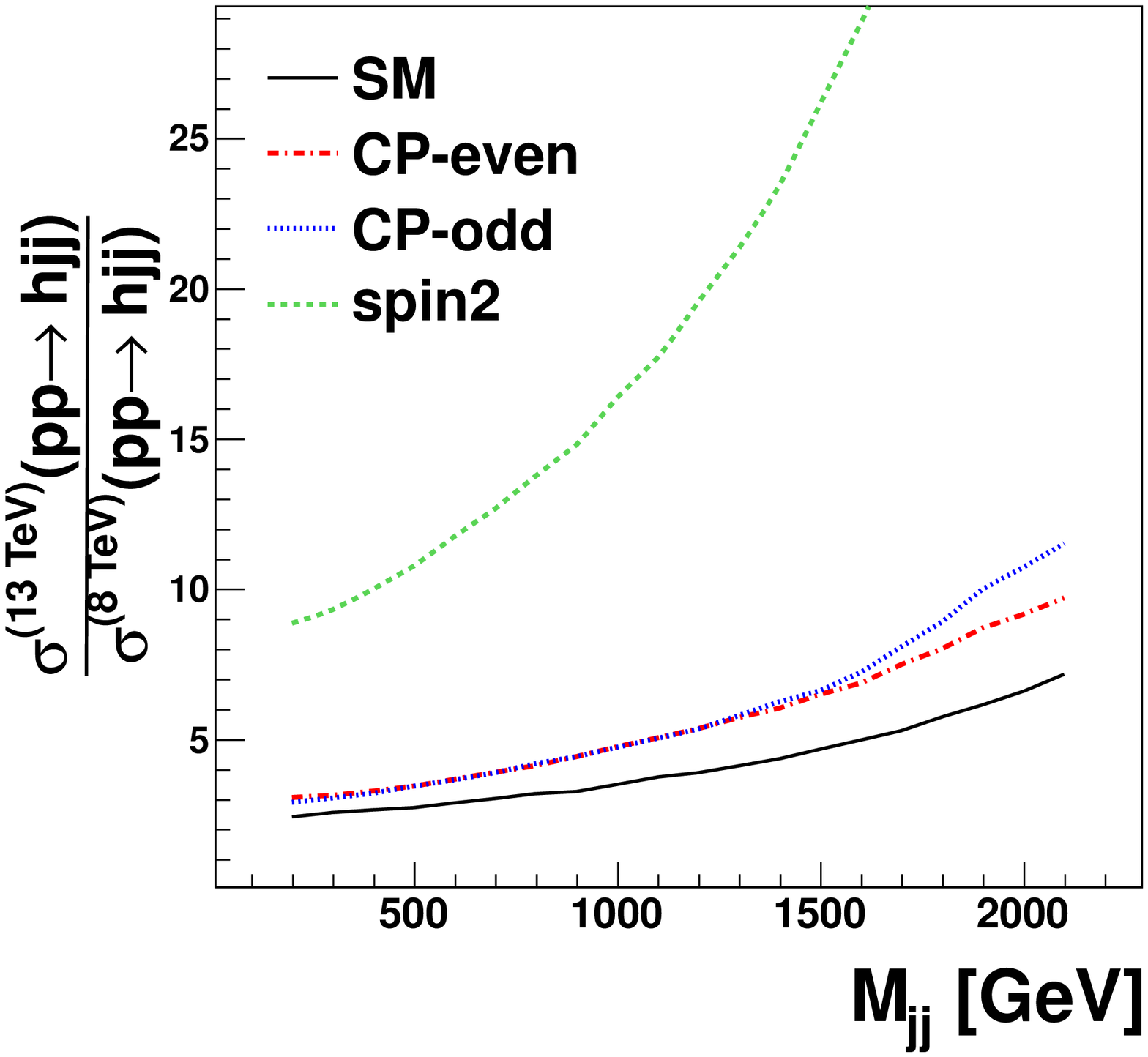}~~
\includegraphics[height=6cm]{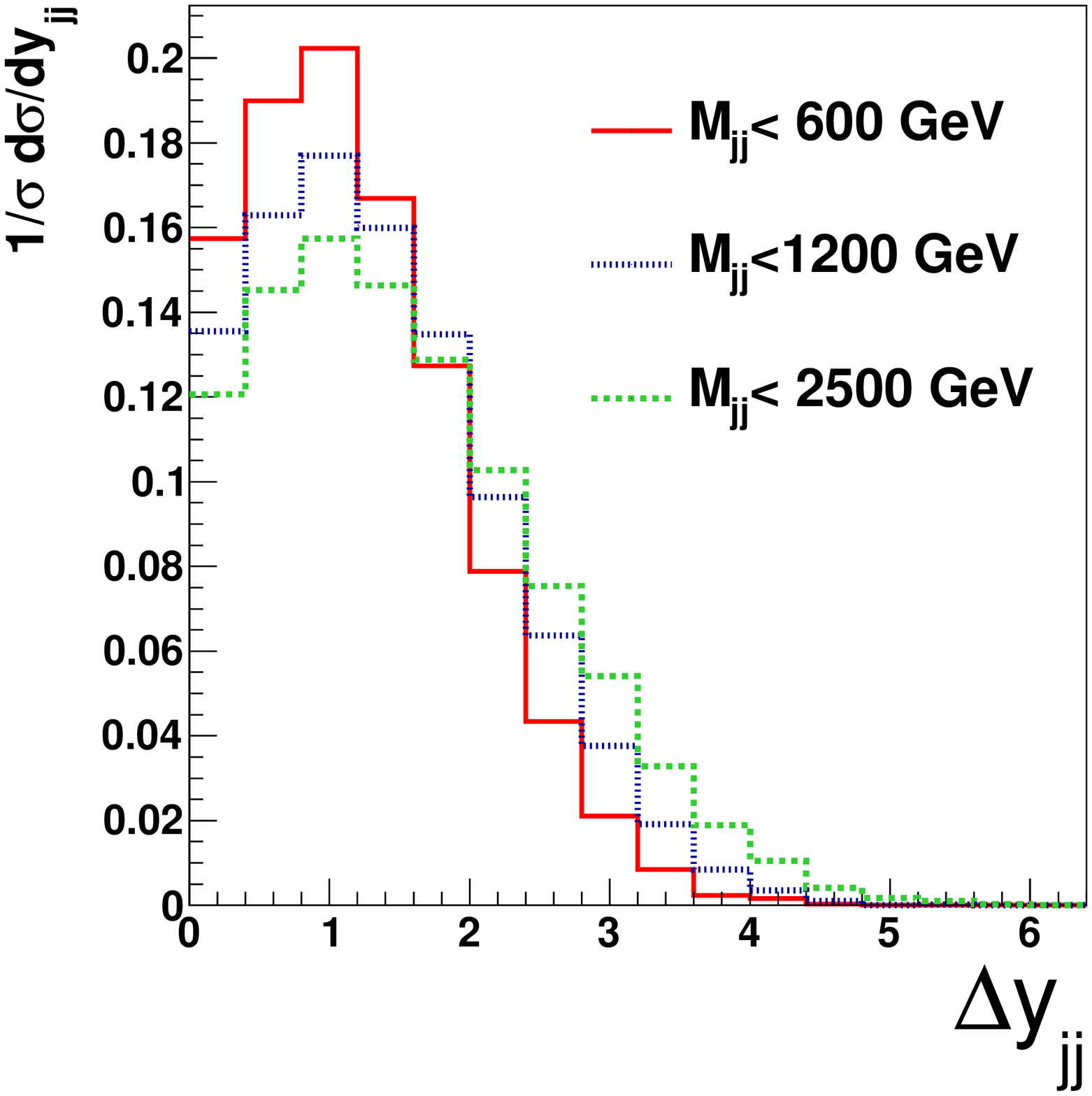} 
\end{center}
\vspace*{-4mm}
\caption{Left: ratio of the cross-sections of the VBF process at 
$\sqrt s= 13\,$TeV to that obtained at $\sqrt s=8\,$TeV as a function
of the lower bound on the invariant mass of the scattered quarks $M_{jj}$; 
results are shown for SM (black solid line) and compared to those of the pure BSM $0^+$ 
(red-dotted-dashed) $0^-$ (blue-dotted) and spin 2 (green-dashed) states. Right: the distributions of the rapidity 
difference of the scattered quarks for three different  values of the upper
bound on $M_{jj}$. In all plots a bound $M_{jj}>200$ GeV  has been set. }

\label{fig:vbfcms}
\vspace*{-3mm}
\end{figure*}

Another discriminating feature is the relative increase of the cross-section
with the energy of the collisions. This feature can be studied when the LHC
begins its higher energy run. Figure~\ref{fig:vbfcms} shows how the ratio of the 
cross-sections at $\sqrt{s}=8\,$TeV and $\sqrt{s}=13$\,TeV for the higher dimension 
operators differ from the SM case as the lower cutoff on the value of the di-parton invariant mass ($M_{jj}$) is changed (x-axis of the plot). 

The advantage of using these
ratios, besides that they are free from many systematics \cite{juan},
 is that they are independent of the values of the couplings 
$\lambda$, $
\lambda^{\prime} $, $f_1$ and $f_2$. This ratio is greater than unity and 
rises with the $M_{jj}$ cut for all the SM and BSM cases under consideration.
For the spin-0 case this ratio is larger for the BSM operators as 
compared to SM. This occurs in spite of the total cross-section (without the 
application of any cuts) rising at the same rate (an increase by a factor of 
$2.3$ from 8 TeV to 13 TeV) with $\sqrt{s}$ for $\Gamma_{\mu\nu}^{\rm SM}$ 
as well as each of the terms in $\Gamma_{\mu\nu}^{\rm BSM}$. 

This result is a mere consequence of the difference in the energy dependence 
of the acceptance for different spin-parity combinations. 

As noted above for spin-2, the increase is much larger $(\approx 7)$ and in fact
the ratios of cross-sections grow much more rapidly for large values of the
$M_{jj}$ cut. This raises a concern about possible  violation of unitarity for
large $M_{jj}$ values. However, there are several ways in which the spin-2 model
can be unitarised. If the unitarisation prescription were to cut off the phase
space concerned with large transverse momentum of the jets, then the $\Delta
y_{jj}$ distributions would shift to larger values simply due to the kinematic
structure of the VBF mechanism. However, in a reasonable model, the unitarisation
may also be implemented such that large values of $M_{jj}$ are cut off.  In this
case, the peak in the $\Delta y_{jj}$ distributions at small values will remain
as shown in the right panel of Fig.~\ref{fig:vbfcms}.

So far, our discussion was independent of the strength of the couplings in the
$HVV$ vertex. While this coupling strength is fixed in the SM, it is not the
case for the BSM couplings and clearly this will affect the actual values of the
cross-sections. Setting $\lambda=0.4$, $\lambda^\prime=0.47$  and $f_1=0.37=f_2$
ensures that $\sigma^{\rm SM}\approx \sigma^{\rm BSM}$. An additional
complication is worth noting. The acceptance for these operators are not the
same in the same regions of phase space and caution is needed when correlating the number
of observed events to a total cross-section. For example, applying an $M_{jj}$
cut removes more events from the additional operators than from SM.  One must be
more careful with the total rate when looking at admixtures of these
operators, which is possible only for the spin-0 case. The value of 
$\sigma^{\rm BSM}$
may increase or decrease depending on the sign, magnitude and on
whether the couplings are real or complex.

\begin{figure*}[t]
\begin{center}
\includegraphics[height=6cm]{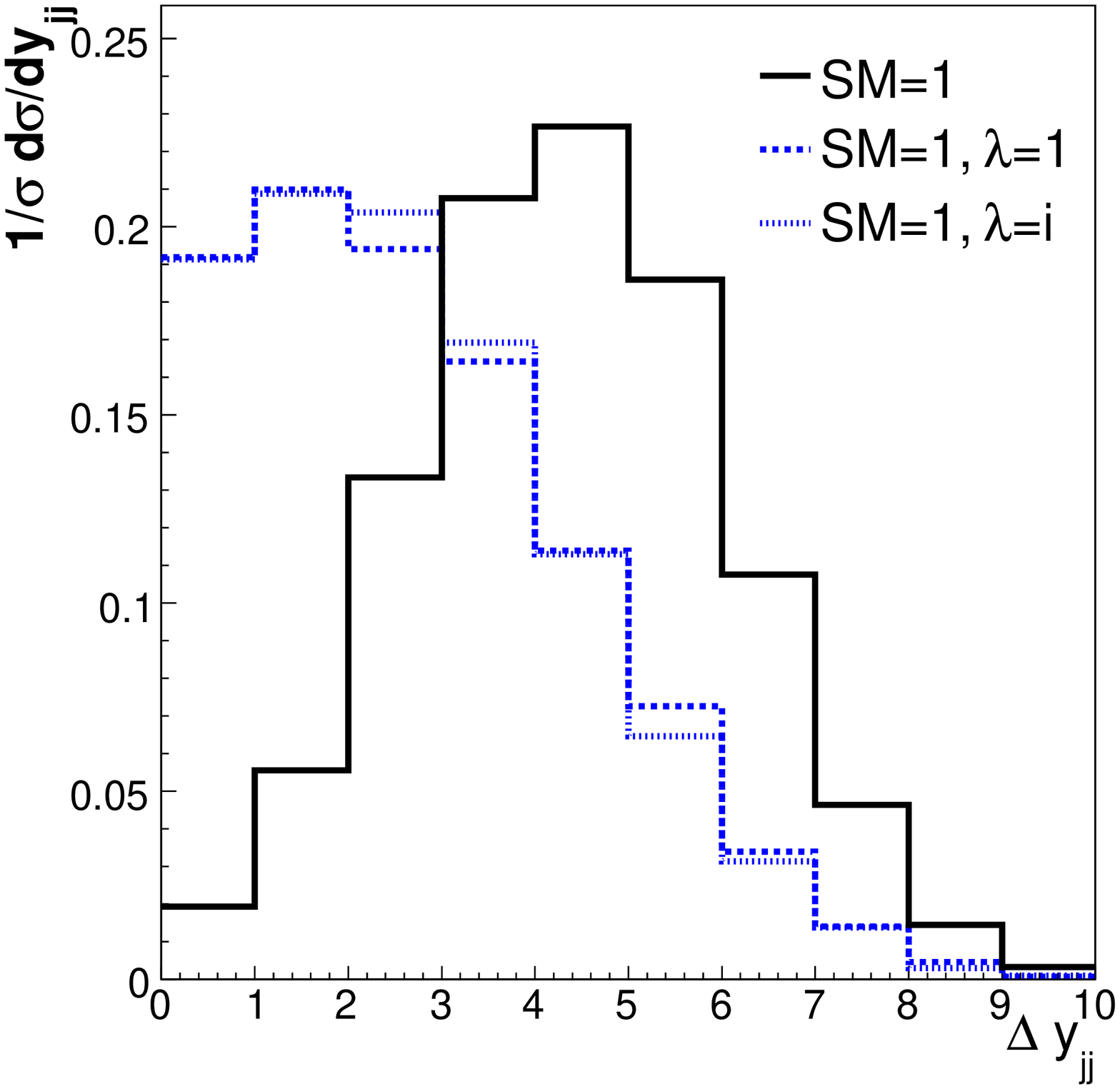}~~~
\includegraphics[height=6cm]{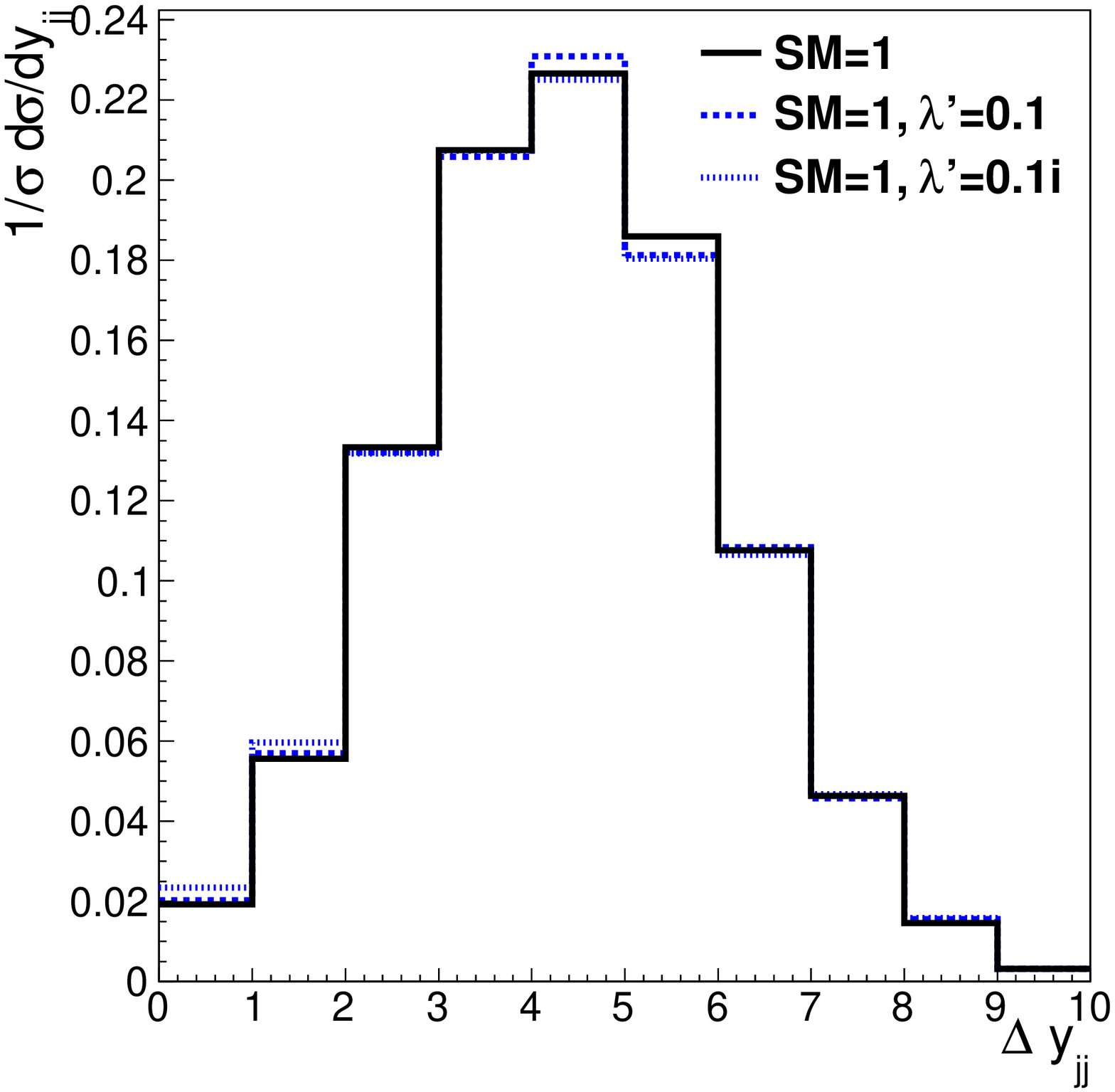}
\end{center}
\vspace*{-3mm}
\caption[]{Normalized distributions of the difference in rapidity between the 
scattered quarks in VBF production: the SM Higgs case (solid 
lines) is compared with the case of BSM admixtures with a $0^+$ term (left) 
and a $0^-$ term (right) as discussed in the text.}
\label{fig:vbf1Db}
\vspace*{-3mm}
\end{figure*}

A cleaner discrimination between the SM and BSM may be possible when the information on rates is supplemented with that on the  kinematic distributions. To this end let us now consider effect
 on these,  of the simultaneous presence of both SM and BSM operators. 
Of course, this  can happen only for spin-0.
In Fig.~\ref{fig:vbf1Db}, the left panel compares the expectation from the
SM (solid line) with an admixture with a BSM $0^+$ scalar with $\lambda=1$
(dashed line) and $\lambda=i$ (dotted line). The right panel shows the
same for the $0^-$ scalar with $\lambda'=0.1\!\cdot\! i$ and  $\lambda'=0.1$. 
Only the strength of the coupling, and not the choice of BSM operators, affect
this distribution. Hence, the plot on the right would look the same if a $0^-$
operator instead of the $0^+$ one were used along with SM. 
 $\Delta y_{jj}$ can therefore effectively  be used
to differentiate a SM from a momentum dependent $HVV$ vertex, for larger values
of these couplings. 
The plot on the left illustrates the fact that for not so small values of 
$\lambda,\lambda^\prime$, it is no longer possible to disentangle the SM Higgs 
boson from admixtures with BSM states using rapidity distributions. However, 
the angular correlation $\Delta\phi_{jj}$
still shows small deviations from the SM case for $\lambda^\prime=0.1$. One
should note that 
the difference in the $\Delta\phi_{jj}$ distributions can be enhanced by looking at events with
large $p_{Tj}> 80$ GeV. A similar observation is valid for comparable values of
$\lambda$.

An important and interesting observation when considering admixtures of the BSM
operators with SM is that for a certain choice of the parameters, SM = 1 and
$\lambda = \lambda^{\prime} = 1$, the $\Delta\phi_{jj}$ distribution becomes
indistinguishable from the SM case. Although this is a special case and the
cross-section is much larger here, it accentuates the importance of the $\Delta
y_{jj}$ distribution and makes it necessary to complement them with $\Delta
\phi_{jj}$ distributions.

\section{Kinematics in the VH mechanism}
\label{sec:vh}

Recently, the invariant mass of the di-boson system in  associated Higgs
production with a vector boson has been suggested as a useful observable to
distinguish the case of a SM Higgs boson from pure BSM  $0^-$ and $2^+$ states
Ref.~\cite{Ellis:2012xd}.  We  briefly survey the effect  of admixtures of the
SM operator with the BSM CP--even and CP--odd operator in the case of the VH
production mechanism.

\begin{figure*}[!h]
\begin{center}
\includegraphics[height=3.8cm]{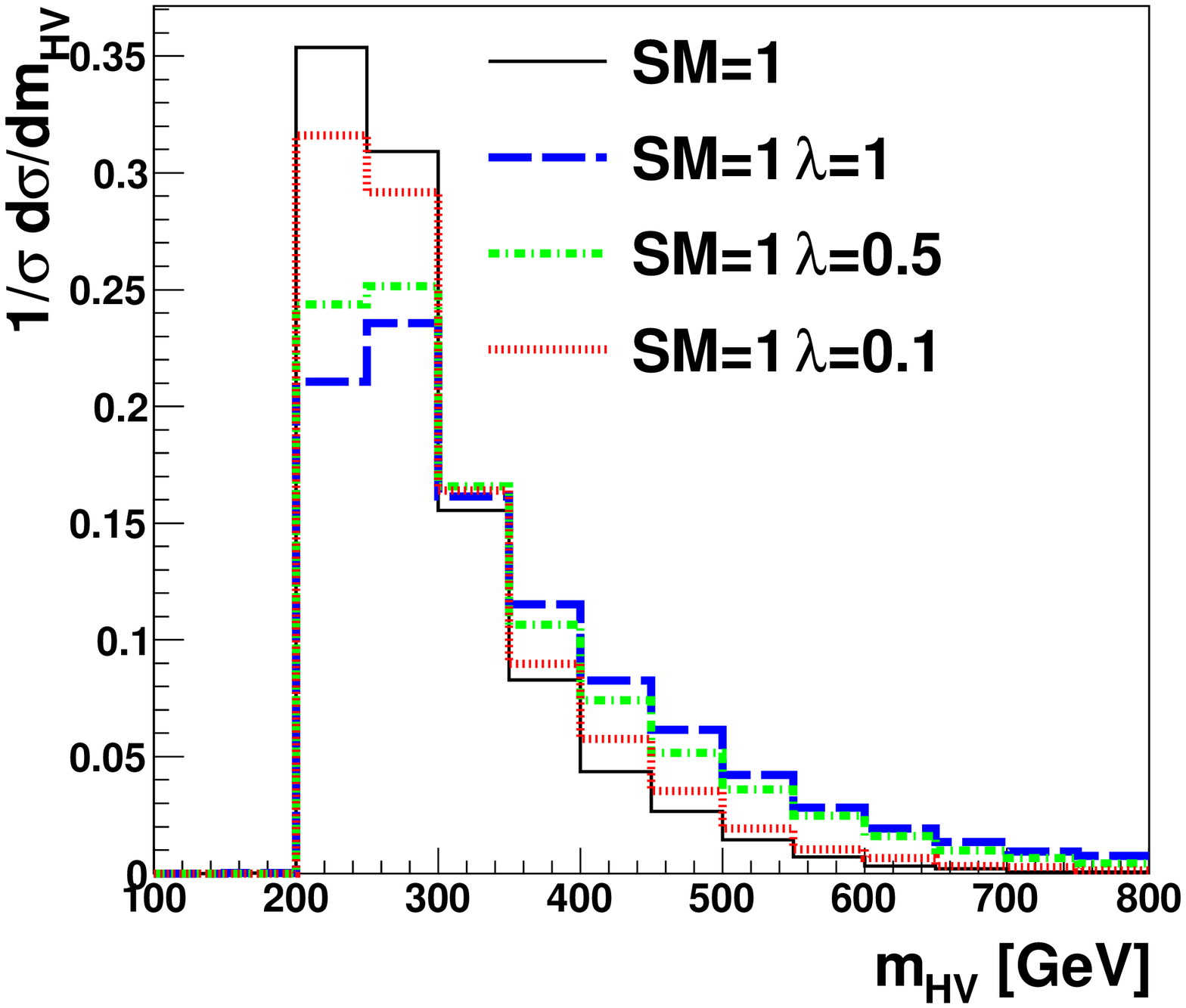} 
\includegraphics[height=3.8cm]{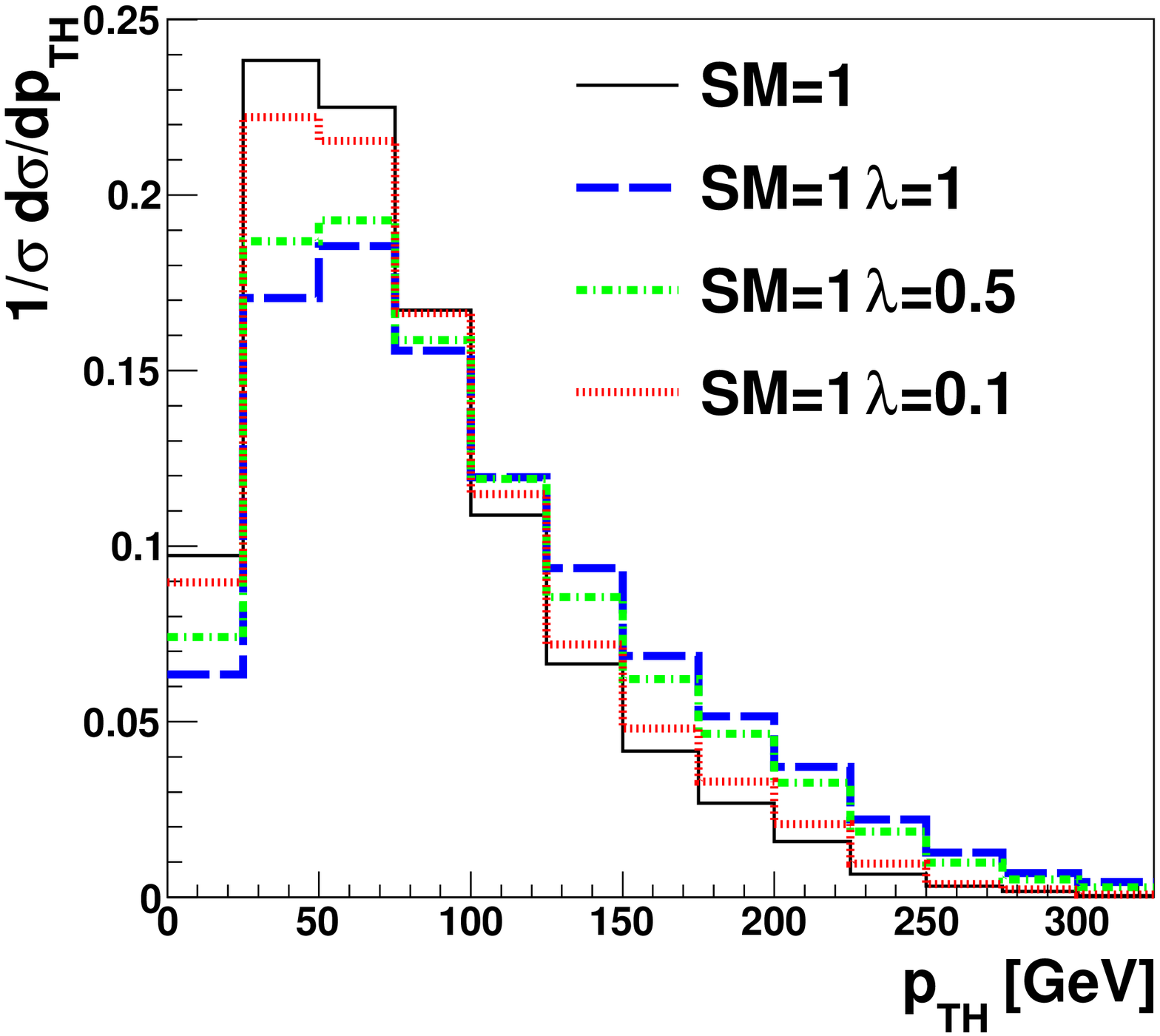}
\includegraphics[height=3.8cm]{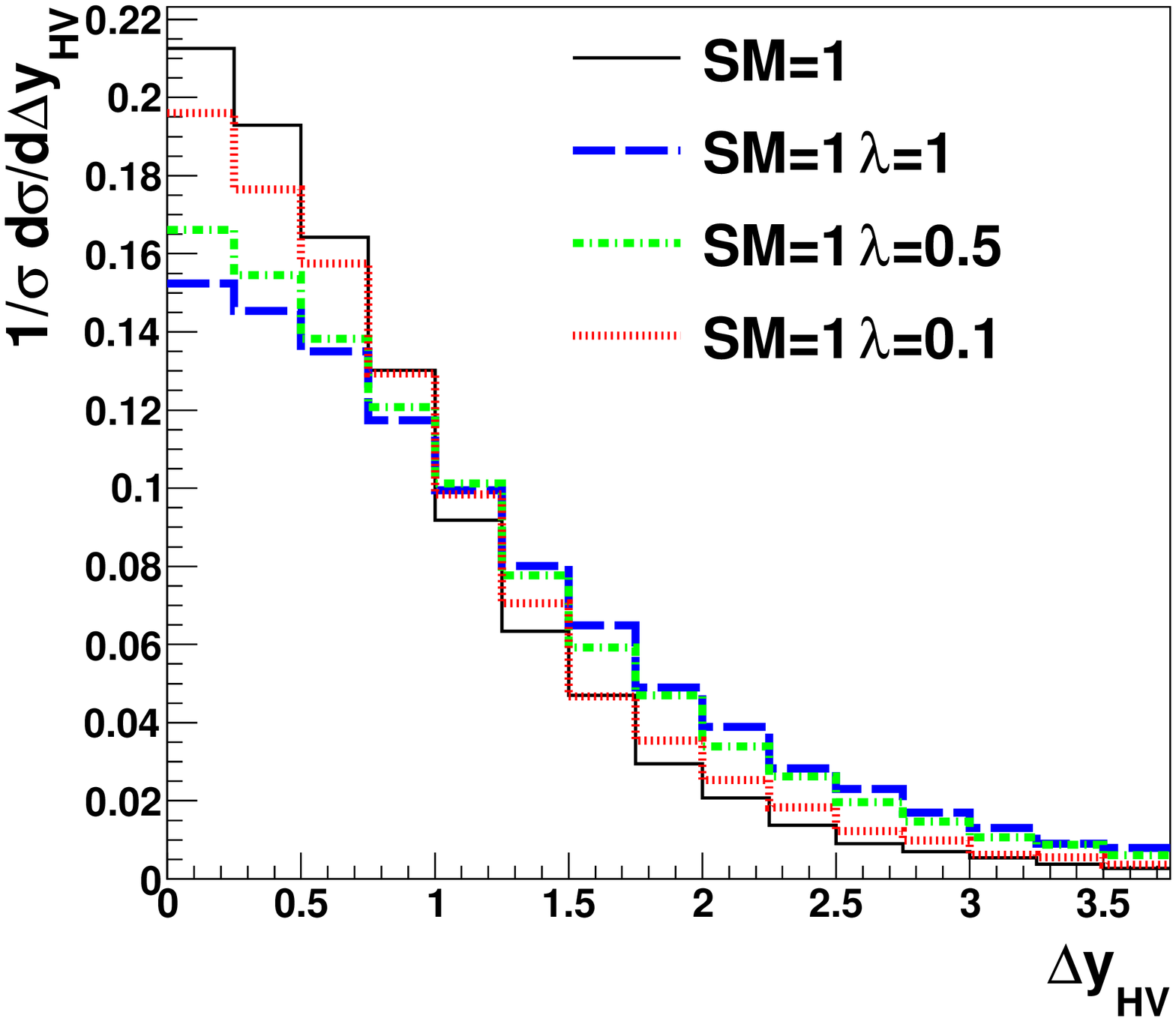}
\end{center}
\caption[]{Invariant mass of the Higgs and the weak boson (left), the transverse momentum of the boson (center) and the  separation in rapidity between the bosons (right). The solid line corresponds to the SM Higgs boson. The blue dashed, green dotted-dashed and red dotted lines correspond to admixtures of the SM Higgs boson with a scalar CP-even state with $\lambda=1, 0.5$ and $0.1$, respectively.   }
\label{fig:vh1}
\end{figure*}

Figure~\ref{fig:vh1} shows the invariant mass of the di-boson system, as well as
the  transverse momentum and the separation in rapidity between the Higgs and
vector bosons. It compares the prediction for the SM Higgs boson and different
admixtures with the CP even operator. With the enhancement of the boson
transverse momentum, the overlap with the BSM electroweak region increases, as
illustrated in Tab.~\ref{tab:BSMEW}.  

We see that along with the invariant mass of the system, the rapidity difference
and the transverse momentum of the bosons are also useful observables to
identify the nature of the $HVV$ interactions. Both the BSM CP-even and CP-odd
operators behave identically and tend to push all three distributions to larger 
values. With diminishing value of the BSM couplings these distributions tend to
become more like the SM distributions, making it harder to determine the BSM couplings  at the
LHC. However unlike the rapidity difference shown in Fig.~\ref{fig:vbf1Db} for
VBF, these observables are more sensitive to the vertex structure.

\section{Conclusions}
\label{sec:conclusions}

Probing the spin and CP structure of the recently discovered boson at the LHC is
of prime importance. In this paper, the sensitivity to physics beyond the SM  via
anomalous Higgs couplings to weak bosons is considered, where contributions from
admixtures of BSM spin-0 and spin-2 states to the $HVV$ coupling are tested.  A
phenomenological survey is performed by exploring various observables in the
electroweak production of the Higgs boson 	with two high $p_T$ jets.
It is found that the kinematics of the tagging jets in the vector boson fusion
mechanism is strongly modified in the presence of momentum dependent anomalous
$HVV$ couplings. In particular,  compared to the SM, the correlation between the
separation in rapidity and the transverse momenta of the scattered quarks varies
significantly in the presence of new physics. The separation in rapidity is
reduced while the transverse momenta becomes significantly larger. In addition,
there is some complementarity between the  behavior of the azimuthal and rapidity
separation between the scattered quarks.  This defines a new corner of the
phase-space that has not been explored by the ATLAS and CMS experiments to test
the spin and CP structure of the Higgs to vector boson coupling.

When this analysis was in the completion stage, Ref.~\cite{Englert:2012xt}
appeared on the archives, wherein the use of the rapidity separation $\Delta
y_{jj}$  to distinguish between various CP and spin states is also advocated. In
our paper, we have in addition investigated the effect of  anomalous couplings
on the correlations between $\sqrt{p_{Tj1}p_{Tj2}}$ and $\Delta y_{jj}$, thereby
uncovering new regions of phase--space for exploiting the VBF production
process. Besides showing that the acceptance of the VBF process to the $M_{jj}$
cut is different for different spin and parity assignments, we have evaluated
how the acceptance (and hence the various kinematical variables)  varies with
the center of mass energy of the proton-proton collision and explored the use of
relative rates between 13\,TeV and 8\,TeV LHC to sharpen up the differences.
Finally, in addition to  the VBF process, we have also considered  associated
Higgs production with vector bosons  and   shown  that  some kinematical
variables such as the invariant mass of the  $VH$ system are sensitive to
anomalous couplings too. \bigskip

{\bf Acknowledgements.} BM  is supported by the DOE Grant No. DE- FG0295-ER40896
and wishes to thank the research Office and the Faculty of Science of the
University of the Witwatersrand. RG wishes to thank the Department of Science
and Technology, Government  of India, for support under grant no.
SR/S2/JCB-64/2007. KM acknowledges CSIR for financial support. AD thanks the
CERN TH unit for the kind hospitality; he is supported by the ERC Advanced
Grant  Higgs@LHC. The authors would like to thank A.~Kruse and Z.~Zhang for the
careful reading of the manuscript. RG and KM would like to acknowledge collaboration with David Miller and Chris White on the
the implementation of FEYNRULES for the anomalous $HVV$ couplings.  


\bibliographystyle{model1-num-names}



\end{document}